\documentclass[11pt,a4paper]{article}
\usepackage{amsmath,amsfonts,amssymb,amscd,graphicx}
\catcode`\@=11
\@addtoreset{equation}{section}

\catcode`\@=12
\newfont{\gotico}{eufm10 scaled\magstephalf}
\newfont{\qvd}{msam10 scaled\magstephalf}
\def\de#1/de#2{\frac{\partial {#1}}{\partial {#2}}}
\def\De#1/de#2{\dfrac{\partial {#1}}{\partial {#2}}}

\begin{document}
%%%%%%%%%%%%%%%%%%%%%%%%%%%%%%%%%%%%%%%%%%%%%%%%%%%%%%%%%%%%%%%%%%%%%%%%%%%%%%%%%%%%%%%%%%%%%%%%%%%
%\baselineskip=2\baselineskip
\vskip-2cm
%%%%%%%%%%%%%%%%%%%%%%%%%%%%%%%%%%%%%%%%%%%%%%%%%%%%%%%%%%%%%%%%%%%%%%%%%%%%%%%%%%%%%%%%%%%%%%%%%%%
\title{Dirac spinors in Bianchi-I $f(R)$-cosmology with torsion}
\author{Stefano Vignolo$^{1}$\footnote{E-mail: vignolo@diptem.unige.it},
Luca Fabbri$^{1,2}$\footnote{E-mail: fabbri@diptem.unige.it}
and Roberto Cianci$^{1}$\footnote{E-mail: cianci@diptem.unige.it}\\
\footnotesize{$^{1}$DIPTEM Sez. Metodi e Modelli Matematici, Universit\`{a} di Genova}\\
\footnotesize{Piazzale Kennedy, Pad. D - 16129 Genova (Italia)}\\
\footnotesize{$^{2}$INFN \& Dipartimento di Fisica, Universit\`{a} di Bologna}\\
\footnotesize{Via Irnerio 46, - 40126 Bologna (Italia)}}
\date{\today}
%%%%%%%%%%%%%%%%%%%%%%%%%%%%%%%%%%%%%%%%%%%%%%%%%%%%%%%%%%%%%%%%%%%%%%%%%%%%%%%%%%%%%%%%%%%%%%%%%%%
\maketitle
%%%%%%%%%%%%%%%%%%%%%%%%%%%%%%%%%%%%%%%%%%%%%%%%%%%%%%%%%%%%%%%%%%%%%%%%%%%%%%%%%%%%%%%%%%%%%%%%%%%
\begin{abstract}
\noindent
We study Dirac spinors in Bianchi type-I cosmological models, within the framework of torsional $f(R)$-gravity. We find four types of results: the resulting dynamic behavior of the universe depends on the particular choice of function $f(R)$; some $f(R)$ models do not isotropize and have no Einstein limit, so that they have no physical significance, whereas for other $f(R)$ models isotropization and Einsteinization occur, and so they are physically acceptable, suggesting that phenomenological arguments may select $f(R)$ models that are physically meaningful; the singularity problem can be avoided, due to the presence of torsion; the general conservation laws holding for $f(R)$-gravity with torsion ensure the preservation of the Hamiltonian constraint, so proving that the initial value problem is well-formulated for these models. 
\end{abstract}
%%%%%%%%%%%%%%%%%%%%%%%%%%%%%%%%%%%%%%%%%%%%%%%%%%%%%%%%%%%%%%%%%%%%%%%%%%%%%%%%%%%%%%%%%%%%%%%%%%
\section{Introduction}
The need to enlarge or revise General Relativity (GR) arises from some open problems in modern physics at both ultraviolet and infrared scales. For instance, at astrophysical and cosmological scales the predictions of GR agree with observations only after the theoretical assumption regarding the existence of the so-called dark matter and dark energy. But at fundamental level, up to now there are no experimental evidences that such unknown forms of matter and energy really exist. This fact together with other shortcomings of Einstein's theory represents signals of a possible breakdown in our understanding of gravity, and the opportunity of developing extended or alternative theories of gravity is to be seriously considered.
\\
Actually, several extensions of GR have been formulated in the last thirty years; among these, $f(R)$-gravity is one of the most direct and simplest: it consists in relaxing the hypothesis that the gravitational Lagrangian be a linear function of the Ricci scalar $R$, thus allowing a dependence on $R$ more general than that of the Einstein-Hilbert action. In the last years, $f(R)$-gravity has received great attention due to the fact that it seems to account for both cosmic speed-up and missing matter at cosmological and astrophysical scales, respectively \cite{n-o,o,m/1}.
\\
At the same time, another way to extend GR consists in considering torsion, including it among the geometrical properties of space-time. Torsion was first introduced in the geometrical background by Cartan; then Sciama and Kibble embodied it into the framework of Einstein gravity following the idea that energy is the source of curvature in the same way in which spin is the source of torsion. The resulting Einstein-Cartan-Sciama-Kibble (ECSK) theory has been the first generalization of GR trying to take the spin of elementary fields into account and it is still one of the most serious attempts in this direction \cite{hehl,m/2}.
\\
According to this line of thinking, $f(R)$-gravity with torsion represents one of the simplest extensions of the ECSK theory, just as purely metric $f(R)$-gravity is with respect to GR. Indeed, the basic idea is again that of replacing the Einstein-Hilbert Lagrangian with a non-linear function. An important consequence of the non-linearity of the gravitational Lagrangian is that we have non-vanishing torsion even without spin, as long as the stress-energy trace is not constant \cite{CCSV1,CCSV2,CCSV3,CV4}. It is known that torsion may give rise to singularity-free and accelerated cosmological models  \cite{SIV}; just arising from the non-linearity of the gravitational Lagrangian function $f(R)$, a propagating torsion could amplify this effect and this point certainly deserves further attention.
\\
Besides, in a recent paper \cite{FV1} we have studied in detail $f(R)$-gravity with torsion in presence of Dirac fields, in order to better exploit the coupling between torsion and spin within the framework of $f(R)$-theories. 
\\
In the present work we explore possible cosmological applications of the theory proposed in \cite{FV1}. More precisely, we study Dirac fields in Bianchi type-I (BI) $f(R)$-cosmological models. BI is the simplest anisotropic space-time that generalizes the spatially flat Friedmann-Lema\"{\i}tre-Robertson-Walker (FLRW) universe. The difference with respect to the FLRW flat model is that in a BI universe the scale factors of spatial directions can differ from each other, thus yielding anisotropy.
\\
Though FLRW models provide very accurate descriptions of the universe observed today, the latter could have undergone an anisotropic phase in its early epoch; for instance, the detected anisotropies in the cosmic microwave background could represent evidences in this direction. On the other hand, non trivial Dirac fields are seen to generate spin-torsion which is incompatible with spatial isotropy \cite{t}: when dealing with Dirac fields in presence of torsion, anisotropic models are then a forced choice. Of course, for consistency with the observed data at present day, the initial anisotropic phase have to undergo a subsequent isotropization process as the Dirac fields energy and spin decrease and universe expands. This dynamical behavior can represent a selection rule for viable functions $f(R)$.
\\
The layout of the paper is the following. In Sections 2, we review some generalities about $f(R)$-gravity with torsion in presence of Dirac fields. In Section 3, we introduce BI cosmological models in torsional $f(R)$-gravity still in presence of Dirac fields and, for completeness, perfect fluid. Following the line traced in previous papers \cite{Saha1,Saha3,Saha2}, we derive a solution of the generalized Dirac equations, holding independently of the assumed model $f(R)$; we integrate the Einstein-like equations in the simplest case $f(R)=R$ and discuss an approximate solution for $f(R)=R+\alpha R^2$. Section 4 is devoted to discuss the field equations in the Einstein frame; we show that, through a conformal transformation from the Jordan to the Einstein frame, we can put the Einstein-like equations in a quadratically integrable form; this result turns out to be useful for a qualitative analysis of the solutions or numerical integration even when explicit calculations for analytical integration are difficult; two examples are given. Finally, in Section 5, we face a crucial aspect in dealing with Einstein-like equations, namely the preservation in time of the Hamiltonian constraint (momentum constraints are here trivial identities); in GR this result is generally ensured by the Bianchi identities and the matter field equations, but here it is not {\it a priori} expected and it has to be proved; anyway, by working out the general conservation laws proved in \cite{FV1}, we show that classical results by Bruhat \cite{yvonne4} apply also in the present case, thus ensuring the required time-preservation of the Hamiltonian constraint. 
%%%%%%%%%%%%%%%%%%%%%%%%%%%%%%%%%%%%%%%%%%%%%%%%%%%%%%%%%%%%%%%%%%%%%%%%%%%%%%%%%%%%%%%%%%%%%%%%%%%
\section{Dirac fields in $f(R)$-gravity with torsion}
In previous papers \cite{CCSV1,CCSV2,CV4,Rubilar}, $f(R)$-gravity with torsion has been formulated within both the metric-affine and the tetrad-affine framework. According to the basic paradigm of $f(R)$-gravity, the gravitational Lagrangian of the theory is assumed to be a real function $f(R)$, where $R$ is here the Ricci curvature scalar written in terms of a metric $g$ and a $g$-compatible connection $\Gamma$, or equivalently a tetrad field $e$ and a spin-connection $\omega$. The pairs $(g,\Gamma)$ and $(e,\omega)$ represent the gravitational dynamical fields of the theory in the metric-affine and tetrad-affine approaches respectively. The field equations turn out to be 
\begin{subequations}
\label{2.1}
\begin{equation}
\label{2.1a}
f'\/(R)R_{ij} -\frac{1}{2}f\/(R)g_{ij}=\Sigma_{ij}
\end{equation}
\begin{equation}
\label{2.1b}
T_{ij}^{\;\;\;h}
=\frac{1}{f'(R)}
\left[\frac{1}{2}\left(\de{f'(R)}/de{x^{p}}+S_{pq}^{\;\;\;q}\right)
\left(\delta^{p}_{j}\delta^{h}_{i}-\delta^{p}_{i}\delta^{h}_{j}\right)
+S_{ij}^{\;\;\;h}\right]
\end{equation}
\end{subequations}
where $R_{ij}$ and $T_{ij}^{\;\;\;h}$ are the Ricci and Torsion tensors, while $\Sigma_{ij}$ and $S_{ij}^{\;\;\;h}$ are the stress-energy and spin density tensors of the matter fields. From \eqref{2.1b}, it is seen that we can have non-vanishing torsion even in absence of spin density.
\\
Making use of the Bianchi identities, it is possible to derive the conservation laws 
\begin{subequations}
\label{2.2}
\begin{equation}
\label{2.2a}
\nabla_{i}\Sigma^{ij}+T_{i}\Sigma^{ij}-\Sigma_{pi}T^{jpi}-\frac{1}{2}S_{sti}R^{stij}=0
\end{equation}
\begin{equation}
\label{2.2b}
\nabla_{h}S^{ijh}+T_{h}S^{ijh}+\Sigma^{ij}-\Sigma^{ji}=0
\end{equation}
\end{subequations}
under which the stress-energy and spin density tensors of the matter fields must undergo once the matter field equations are assigned \cite{FV1}. In equations \eqref{2.2} the symbols $\nabla_i$ denote covariant derivative with respect to the dynamical connection $\Gamma$ while $R^{stij}$ is the curvature tensor of $\Gamma$. Indices are lowered and raised by the metric $g_{ij}$.
\\
In this paper we deal with $f(R)$-gravity coupled with Dirac fields and possibly a perfect fluid (without spin). Denoting by $\gamma^\mu$ ($\mu=0,1,2,3$) Dirac matrices, we introduce the notation $\Gamma^i = e^i_\mu\gamma^\mu$ where $e^\mu_i$ ($e^\mu_{i}e^i_\nu=\delta^\mu_\nu$ and $e^i_{\mu}e^\mu_j=\delta^i_j$) indicate a tetrad field associated with a metric $g_{ij}$. Moreover, setting $S_{\mu\nu}:= \frac{1}{8}[\gamma_\mu,\gamma_\nu]$, we denote the covariant derivative of the Dirac field $\psi$ by $D_i\psi = \de\psi/de{x^i} + \omega_i^{\;\;\mu\nu}S_{\mu\nu}\psi\/$ and $D_i\bar\psi = \de{\bar\psi}/de{x^i} - \bar\psi\omega_i^{\;\;\mu\nu}S_{\mu\nu}\/$, where $\omega_i^{\;\;\mu\nu}$ is a spin connection. Equivalently, we can put $D_i\psi = \de\psi/de{x^i} - \Omega_i\psi$ and $D_i\bar\psi = \de{\bar\psi}/de{x^i} + \bar{\psi}\Omega_i$ where
\begin{equation}\label{2.3}
\Omega_i := - \frac{1}{4}g_{jh}\left(\Gamma_{ik}^{\;\;\;j} - e^j_\mu\partial_i\/e^\mu_k \right)\Gamma^h\Gamma^k
\end{equation}
$\Gamma_{ik}^{\;\;\;j}$ being the coefficients of a linear connection $\Gamma$. The relation between linear and spin connection is given by
\begin{equation}\label{2.4}
\Gamma_{ij}^{\;\;\;h} = \omega_{i\;\;\;\nu}^{\;\;\mu}e_\mu^h\/e^\nu_j + e^{h}_{\mu}\partial_{i}e^{\mu}_{j}
\end{equation}
The stress-energy tensors of the matter fields are then described as
\begin{subequations}\label{2.5}
\begin{equation}\label{2.5a}
\Sigma^D_{ij} := \frac{i}{4}\/\left( \bar\psi\Gamma_{i}{D}_{j}\psi - {D}_{j}\bar{\psi}\Gamma_{i}\psi \right)
\end{equation}
and 
\begin{equation}\label{2.5b}
\Sigma^F_{ij}:= (\rho +p)\/U_iU_j -pg_{ij}
\end{equation}
while the spin density tensor is expressed as
\begin{equation}\label{2.5c}
S_{ij}^{\;\;\;h}=\frac{i}{2}\bar\psi\left\{\Gamma^{h},S_{ij}\right\}\psi
\equiv-\frac{1}{4}\eta^{\mu\sigma}\epsilon_{\sigma\nu\lambda\tau}
\left(\bar{\psi}\gamma_{5}\gamma^{\tau}\psi\right)e^{h}_{\mu}e^{\nu}_{i}e^{\lambda}_{j}
\end{equation}
\end{subequations}
In equations \eqref{2.5}, $\rho$, $p$ and $U_i$ denote respectively the matter-energy density, the pressure and the four velocity of the fluid, $S_{ij}:= \frac{1}{8}[\Gamma_i,\Gamma_j]$ and $\eta^{\mu\sigma}$ is the Minkowskian metric with signature $(1,-1,-1,-1)$. 
\\
Following previous works \cite{CCSV1,CCSV2,CCSV3,CV4,FV1}, we suppose that the trace of equations \eqref{2.1a}
\begin{equation}\label{2.6}
f'(R)R -2f(R)=\Sigma
\end{equation}
gives rise to an invertible relation between the Ricci scalar curvature $R$ and the trace $\Sigma$ of the stress-energy tensor. Also, we assume that $f(R)\not = kR^2$ (the case $f(R)=kR^2$ is only compatible with the condition $\Sigma=0$). Under the assumed conditions, from equation \eqref{2.6} it is possible to derive the expression of $R$ as function of $\Sigma$, namely $R=F(\Sigma)$. After that, introducing the scalar field 
\begin{equation}\label{2.7}
\varphi := f'\/(F\/(\Sigma))
\end{equation}
and the effective potential
\begin{equation}\label{2.8}
V\/(\varphi):= \frac{1}{4}\left[ \varphi F^{-1}\/((f')^{-1}\/(\varphi))+ \varphi^2\/(f')^{-1}\/(\varphi)\right]
\end{equation}
as well as separating the Levi--Civita contributions from the torsional ones, we can express the field equations \eqref{2.1a} in the Einstein-like form
\begin{equation}\label{2.9}
\begin{split}
\tilde{R}_{ij} -\frac{1}{2}\tilde{R}g_{ij}= \frac{1}{\varphi}\Sigma^F_{ij} + \frac{1}{\varphi}\Sigma^D_{ij} + \frac{1}{\varphi^2}\left( - \frac{3}{2}\de\varphi/de{x^i}\de\varphi/de{x^j} + \varphi\tilde{\nabla}_{j}\de\varphi/de{x^i} + \frac{3}{4}\de\varphi/de{x^h}\de\varphi/de{x^k}g^{hk}g_{ij} \right. \\
\left. - \varphi\tilde{\nabla}^h\de\varphi/de{x^h}g_{ij} - V\/(\varphi)g_{ij} \right) + \tilde{\nabla}_h\hat{S}_{ji}^{\;\;\;h} + \hat{S}_{hi}^{\;\;\;p}\hat{S}_{jp}^{\;\;\;h} - \frac{1}{2}\hat{S}_{hq}^{\;\;\;p}\hat{S}_{\;\;p}^{q\;\;\;h}g_{ij}
\end{split}
\end{equation}
where $\tilde{R}_{ij}$, $\tilde R$ and $\tilde{\nabla}_i$ denote respectively the Ricci tensor, the Ricci scalar curvature and the covariant derivative of the Levi--Civita connection, and $\hat{S}_{ij}^{\;\;\;h}:=-\frac{1}{2\varphi}S_{ij}^{\;\;\;h}\/$.
\\
In addition to this, the generalized Dirac equations for the spinor field are 
\begin{equation}\label{2.10}
i\Gamma^{h}D_{h}\psi + \frac{i}{2}T_h\Gamma^h\psi- m\psi=0
\end{equation}
where $T_h :=T_{hj}^{\;\;\;j}$ is the torsion vector. As it has been proved in \cite{FV1}, equations \eqref{2.10} imply automatically the validity of the conservation laws
\begin{subequations}\label{2.11}
\begin{equation}\label{2.11a}
\nabla_{i}\Sigma^{D\/{ij}}+T_{i}\Sigma^{D\/{ij}}-\Sigma^D_{\/pi}T^{jpi}-\frac{1}{2}S_{sti}R^{stij}=0
\end{equation}
\begin{equation}\label{2.11b}
\nabla_{h}S^{ijh}+T_{h}S^{ijh}+\Sigma^{D\/{ij}}-\Sigma^{D\/{ji}}=0
\end{equation}
\end{subequations}
Moreover, still in \cite{FV1}, it has been shown that equations \eqref{2.11b} are equivalent to the antisymmetric part of the Einstein-like equations \eqref{2.9}. From this, we get two results: on one hand, we have that the significant part of the Einstein-like field equations \eqref{2.9} is the symmetric one; on the other hand, comparing equations \eqref{2.11a} with equations \eqref{2.2a}, we derive the conservation law for the perfect fluid. Indeed, introducing the contorsion tensor 
\begin{equation}\label{2.12}
K_{ij}^{\;\;\;h}
=\frac{1}{2}\left(-T_{ij}^{\;\;\;h}+T_{j\;\;\;i}^{\;\;h}-T^{h}_{\;\;ij}\right)
\end{equation}
and expressing the dynamical connection as the sum
\begin{equation}\label{2.13}
\Gamma_{ij}^{\;\;\;h}=\tilde{\Gamma}_{ij}^{\;\;\;h}-K_{ij}^{\;\;\;h}
\end{equation}
where $\tilde{\Gamma}_{ij}^{\;\;\;h}$ are the coefficients of the Levi--Civita connection, a direct comparison between equations \eqref{2.11a} with equations \eqref{2.2a} yields the identity
\begin{equation}\label{2.14}
\begin{split}
\nabla_{i}\Sigma^{F\/{ij}}+T_{i}\Sigma^{F\/{ij}}-\Sigma^F_{\/pi}T^{jpi}=
\tilde{\nabla}_i\Sigma^{F\/{ij}} - K_{ih}^{\;\;\;i}\Sigma^{F\/{hj}} - K_{ih}^{\;\;\;j}\Sigma^{F\/{ih}}\\ +T_{h}\Sigma^{F\/{hj}}-T^{j}_{\;\;ih}\Sigma^{F\/{ih}} =
\tilde{\nabla}_i\Sigma^{F\/{ij}} =0
\end{split}
\end{equation}
which is the same conservation laws holding in general relativity.
\\
To conclude this preliminary section, we notice that the symmetrized part of the Einstein-like equations \eqref{2.9} as well as the Dirac equations \eqref{2.10} can be worked out as in \cite{FV1} assuming the final form
\begin{equation}\label{2.15}
\begin{split}
\tilde{R}_{ij} -\frac{1}{2}\tilde{R}g_{ij}= \frac{1}{\varphi}\Sigma^F_{ij} + \frac{1}{\varphi}\tilde{\Sigma}^D_{ij}
+ \frac{1}{\varphi^2}\left( - \frac{3}{2}\de\varphi/de{x^i}\de\varphi/de{x^j} + \varphi\tilde{\nabla}_{j}\de\varphi/de{x^i} + 
\frac{3}{4}\de\varphi/de{x^h}\de\varphi/de{x^k}g^{hk}g_{ij} \right. \\
\left. - \varphi\tilde{\nabla}^h\de\varphi/de{x^h}g_{ij} - V\/(\varphi)g_{ij} \right) + \frac{3}{64\varphi^2}(\bar{\psi}\gamma_5\gamma^\tau\psi)(\bar{\psi}\gamma_5\gamma_\tau\psi)g_{ij}
\end{split}
\end{equation}
and 
\begin{equation}\label{2.16}
i\Gamma^{h}\tilde{D}_{h}\psi
-\frac{1}{\varphi}\frac{3}{16}\left[(\bar{\psi}\psi)
+i(i\bar{\psi}\gamma_5\psi)\gamma_5\right]\psi-m\psi=0
\end{equation}
where
\begin{equation}\label{2.17}
\tilde{\Sigma}^D_{ij} := \frac{i}{4}\/\left[ \bar\psi\Gamma_{(i}\tilde{D}_{j)}\psi - \left(\tilde{D}_{(j}\bar\psi\right)\Gamma_{i)}\psi \right]
\end{equation}
$\tilde{D}_i$ denoting covariant derivative with respect to the Levi--Civita connection.
%%%%%%%%%%%%%%%%%%%%%%%%%%%%%%%%%%%%%%%%%%%%%%%%%%%%%%%%%%%%%%%%%%%%%%%%%%%%%%%%%%%%%%%%%%%%%%%%%%
\section{Bianchi-I cosmological models}
Let us consider a Bianchi type I metric of the form
\begin{equation}\label{3.1}
ds^2 = dt^2 - a^2(t)\,dx^2 - b^2(t)\,dy^2 - c^2(t)\,dz^2
\end{equation}
The tetrad field associated with the metric \eqref{3.1} is given by
\begin{equation}\label{3.2}
e^\mu_0=\delta^\mu_0, \quad e^\mu_1 = a(t)\/\delta^\mu_1, \quad e^\mu_2 = b(t)\/\delta^\mu_2, \quad e^\mu_3 = c(t)\/\delta^\mu_3 \qquad \mu =0,1,2,3
\end{equation}
The inverse relations of \eqref{3.3} are expressed as
\begin{equation}\label{3.3}
e^0_\mu = \delta^0_\mu, \quad e^1_\mu = \frac{1}{a(t)}\delta^1_\mu, \quad e^2_\mu = \frac{1}{b(t)}\delta^2_\mu, \quad e^3_\mu = \frac{1}{c(t)}\delta^3_\mu \qquad \mu =0,1,2,3
\end{equation}
The non-trivial Christoffel symbols associated to the metric \eqref{3.1} are
\begin{equation}\label{3.4}
\begin{split}
\tilde{\Gamma}_{10}^{\;\;\;1}= \frac{\dot a}{a}, \quad \tilde{\Gamma}_{20}^{\;\;\;2}= \frac{\dot b}{b}, \quad \tilde{\Gamma}_{30}^{\;\;\;3}= \frac{\dot c}{c}\\
\tilde{\Gamma}_{11}^{\;\;\;0}= a{\dot a}, \quad \tilde{\Gamma}_{22}^{\;\;\;0}= b{\dot b}, \quad \tilde{\Gamma}_{33}^{\;\;\;0}= c{\dot c}
\end{split}
\end{equation}
In this case, the matrices $\Gamma^i = e^i_\mu\gamma^\mu$ assume the explicit form
\begin{equation}\label{3.5}
\Gamma^0 = \gamma^0,\quad \Gamma^1 = \frac{1}{a(t)}\gamma^1, \quad \Gamma^2 = \frac{1}{b(t)}\gamma^2, \quad \Gamma^3 = \frac{1}{c(t)}\gamma^3 
\end{equation}
The spinorial covariant derivative induced by the Levi--Civita connection is
\begin{equation}\label{3.7}
\tilde{D}_i\psi = \partial_i\psi - \tilde{\Omega}_i\psi, \qquad \tilde{D}_i\bar\psi = \partial_i\bar\psi + \bar{\psi}\tilde{\Omega}_i
\end{equation}
where the spinor connection coefficients $\tilde{\Omega}_i$ are given by
\begin{equation}\label{3.8}
\tilde{\Omega}_0=0, \quad \tilde{\Omega}_1=\frac{1}{2}{\dot a}\gamma^1\gamma^0, \quad \tilde{\Omega}_2=\frac{1}{2}{\dot b}\gamma^2\gamma^0, \quad \tilde{\Omega}_3=\frac{1}{2}{\dot c}\gamma^3\gamma^0
\end{equation} 
Taking equations \eqref{3.7} and \eqref{3.8} into account, it is easily seen that the Dirac equations \eqref{2.16} assume the form
\begin{subequations}\label{3.9}
\begin{equation}\label{3.9a}
\dot\psi + \frac{\dot\tau}{2\tau}\psi + im\gamma^0\psi -
\frac{3i}{16\varphi}\/\left[ (\bar\psi\psi)\gamma^0 +i\/(i\bar\psi\gamma^5\psi)\gamma^0\gamma^5 \right]\psi =0
\end{equation}
\begin{equation}\label{3.9b}
\dot{\bar\psi} + \frac{\dot\tau}{2\tau}\bar\psi - im\bar{\psi}\gamma^0 + \frac{3i}{16\varphi}\bar\psi\/\left[ (\bar\psi\psi)\gamma^0 +i\/(i\bar\psi\gamma^5\psi)\gamma^5\gamma^0 \right] =0
\end{equation}
\end{subequations}
where, borrowing from \cite{Saha1,Saha2}, we have defined $\tau := abc$. Analogously, evaluating the Einstein-like equations \eqref{2.15} for the metric \eqref{3.1}, we get 
\begin{subequations}\label{3.10}
\begin{equation}\label{3.10a}
\begin{split}
\frac{\dot a}{a}\frac{\dot b}{b} + \frac{\dot b}{b}\frac{\dot c}{c} + \frac{\dot a}{a}\frac{\dot c}{c} = \frac{\rho}{\varphi} +
\frac{1}{2\varphi}m\bar\psi\psi - \frac{3}{64\varphi^2}(\bar{\psi}\gamma_5\gamma^\tau\psi)(\bar{\psi}\gamma_5\gamma_\tau\psi)+\\ 
+\frac{1}{\varphi^2}\left[- \frac{3}{4}{\dot\varphi}^2 - \varphi\dot\varphi\frac{\dot\tau}{\tau} - V(\varphi)\right] 
\end{split}
\end{equation}
\begin{equation}\label{3.10b}
\begin{split}
\frac{\ddot b}{b} + \frac{\ddot c}{c} + \frac{\dot b}{b}\frac{\dot c}{c} = - \frac{p}{\varphi} + 
\frac{1}{\varphi^2}\left[\varphi\dot\varphi\frac{\dot a}{a}+\frac{3}{4}{\dot\varphi}^2 -\varphi\left( \ddot\varphi + \frac{\dot\tau}{\tau}\dot\varphi \right) - V(\varphi)\right]+\\ +\frac{3}{64\varphi^2}(\bar{\psi}\gamma_5\gamma^\tau\psi)(\bar{\psi}\gamma_5\gamma_\tau\psi)
\end{split}
\end{equation}
\begin{equation}\label{3.10c}
\begin{split}
\frac{\ddot a}{a} + \frac{\ddot c}{c} + \frac{\dot a}{a}\frac{\dot c}{c} = - \frac{p}{\varphi} + 
\frac{1}{\varphi^2}\left[\varphi\dot\varphi\frac{\dot b}{b} + \frac{3}{4}{\dot\varphi}^2 -\varphi\left( \ddot\varphi + \frac{\dot\tau}{\tau}\dot\varphi \right) - V(\varphi)\right]+\\
+\frac{3}{64\varphi^2}(\bar{\psi}\gamma_5\gamma^\tau\psi)(\bar{\psi}\gamma_5\gamma_\tau\psi)
\end{split}
\end{equation}
\begin{equation}\label{3.10d}
\begin{split}
\frac{\ddot a}{a} + \frac{\ddot b}{b} + \frac{\dot a}{a}\frac{\dot b}{b} = - \frac{p}{\varphi} + 
\frac{1}{\varphi^2}\left[\varphi\dot\varphi\frac{\dot c}{c} + \frac{3}{4}{\dot\varphi}^2 -\varphi\left( \ddot\varphi + \frac{\dot\tau}{\tau}\dot\varphi \right) - V(\varphi)\right]+\\
+\frac{3}{64\varphi^2}(\bar{\psi}\gamma_5\gamma^\tau\psi)(\bar{\psi}\gamma_5\gamma_\tau\psi)
\end{split}
\end{equation}
\end{subequations}
together with the conditions
\begin{subequations}
\label{3.11}
\begin{equation}\label{3.11a}
\tilde{\Sigma}^D_{12}=0\quad \Rightarrow \quad a\/\dot{b} - b\/\dot{a}=0 \quad \cup \quad \bar\psi\gamma^5\gamma^3\psi =0
\end{equation}
\begin{equation}\label{3.11b}
\tilde{\Sigma}^D_{23}=0\quad \Rightarrow \quad c\/\dot{b} - b\/\dot{c}=0 \quad \cup \quad \bar\psi\gamma^5\gamma^1\psi =0
\end{equation}
\begin{equation}\label{3.11c}
\tilde{\Sigma}^D_{13}=0\quad \Rightarrow \quad a\/\dot{c} - c\/\dot{a}=0 \quad \cup \quad \bar\psi\gamma^5\gamma^2\psi =0
\end{equation}
\end{subequations}
The equations $\tilde{\Sigma}^D_{0A}=0$ ($A=1,2,3$) are automatically satisfied identities. Finally, the conservation law \eqref{2.14} together with an equation of state of the kind $p=\lambda\rho$ ($\lambda\in [0,1[$) yield the last equation
\begin{equation}\label{3.12}
\dot\rho + \frac{\dot\tau}{\tau}(1+\lambda)\rho =0
\end{equation}
which completes the whole set of field equations. The general solution of \eqref{3.12} is given by
\begin{equation}\label{3.12bis}
\rho = \rho_0\tau^{-(1+\lambda)} \qquad \rho_0 = {\rm constant}
\end{equation}
\\
Conditions \eqref{3.11} are constraints imposed on the metric or on the Dirac field. We see that there are in general three ways of satisfying these conditions: one is to impose constraints of purely geometrical origin by insisting that $a\dot{b}-b\dot{a}=0$, $a\dot{c}-c\dot{a}=0$, $c\dot{b}-b\dot{c}=0$ giving an isotropic universe filled with fermionic matter fields, which is problematic due to the fact that it is known that Dirac fields do not undergo the cosmological principle \cite{t}; another is to impose constraints of purely material origin by insisting that $\bar\psi\gamma^5\gamma^1\psi=0$, $\bar\psi\gamma^5\gamma^2\psi=0$, $\bar\psi\gamma^5\gamma^3\psi=0$ giving an anisotropic universe without fermionic torsional interactions, which we regard as unsatisfactory because if Dirac fields are absent then it is not clear what may then justify anisotropies; the last situation would be of both geometrical and material origin by insisting that for instance $a\dot{b}-b\dot{a}=0$ with $\bar\psi\gamma^5\gamma^1\psi=0$, $\bar\psi\gamma^5\gamma^2\psi=0$ giving a partial isotropy for only two axes with the corresponding two components of the spin vector vanishing, describing a universe shaped as an ellipsoid of rotation about the only axis along which the spin vector does not vanish. Notice that by insisting on the proportionality between two couples of axes we inevitably get the total isotropy of the $3$-dimensional space. Therefore, the situation in which we have $a=b$ with $\bar\psi\gamma^5\gamma^1\psi=\bar\psi\gamma^5\gamma^2\psi=0$ is the only one that we believe to be entirely satisfactory, and from now on we shall work in this situation.
\\
Here, the Dirac and Einstein-like equations \eqref{3.9} and \eqref{3.10} can be worked out as in \cite{Saha1,Saha2}: for instance, through suitable combinations of \eqref{3.10} we obtain the equations
\begin{subequations}\label{3.13}
\begin{equation}\label{3.13a}
\frac{d}{dt}(\psi^{\dagger}\psi\tau)=0
\end{equation}
\begin{equation}\label{3.13b}
\frac{d}{dt}(\bar\psi\psi\tau)
-\frac{3}{8\varphi}(i\bar\psi\gamma^5\psi)(\psi^{\dagger}\gamma^5\psi\tau)=0
\end{equation}
\begin{equation}\label{3.13c}
\frac{d}{dt}(i\bar\psi\gamma^5\psi\tau)
+\left[2m+\frac{3}{8\varphi}(\bar\psi\psi)\right](\psi^{\dagger}\gamma^5\psi\tau)=0
\end{equation}
\begin{equation}\label{3.13d}
\frac{d}{dt}(\psi^{\dagger}\gamma^5\psi\tau)-2m(i\bar\psi\gamma^5\psi\tau)=0
\end{equation}
\end{subequations} 
From equations \eqref{3.13} it is easy to deduce that
\begin{subequations}
\label{3.13bis}
\begin{equation}\label{3.13abis}
(\bar\psi\gamma^5\gamma^3\psi)^2=(\bar\psi\psi)^2 + (i\bar\psi\gamma^5\psi)^2 + (\bar\psi\gamma^5\gamma^0\psi)^2 = \frac{C^2}{\tau^2}
\end{equation}
\begin{equation}\label{3.13bbis}
(\psi^{\dagger}\psi)^2=\frac{K^2}{\tau^2}
\end{equation}
\end{subequations} 
with $C$ and $K$ constants. We notice that in this special case the theory has an additional discrete symmetry given by the fact that under the discrete transformation $\psi \rightarrow \gamma^5\gamma^0\gamma^1\psi$ all field equations are invariant; this implies that in the Dirac equation the total number of $4$ components is in this case reduced to $2$ components alone; however $2$ components with complex values are equivalent to $4$ components of real values: so the $4$ equations for real fields given in \eqref{3.13} are the system of field equations we will have to solve. The compatibility with all constraints allows only three classes of spinors, each of which has a general member written in the following form
\begin{eqnarray}
\nonumber
\label{generalspinor}
&\psi=\frac{1}{\sqrt{2\tau}}\left(\begin{tabular}{c}
$\sqrt{K-C}\cos{\zeta_{1}}e^{i\theta_{1}}$\\
$\sqrt{K+C}\cos{\zeta_{2}}e^{i\vartheta_{1}}$\\
$\sqrt{K-C}\sin{\zeta_{1}}e^{i\vartheta_{2}}$\\
$\sqrt{K+C}\sin{\zeta_{2}}e^{i\theta_{2}}$
\end{tabular}\right)
\end{eqnarray}
with constraints $\tan{\zeta_{1}}\tan{\zeta_{2}}=(-1)^{n+1}$ and $\theta_{1}+\theta_{2}-\vartheta_{1}-\vartheta_{2}=\pi n$ for any $n$ integer, and also
\begin{eqnarray}
\label{restrictedspinor1}
&\psi=\frac{1}{\sqrt{2\tau}}\left(\begin{tabular}{c}
$\sqrt{K-C}\cos{\zeta_{1}}e^{i\theta_{1}}$\\
$0$\\
$0$\\
$\sqrt{K+C}\sin{\zeta_{2}}e^{i\theta_{2}}$
\end{tabular}\right)
\end{eqnarray}
and
\begin{eqnarray}
\label{restrictedspinor2}
&\psi=\frac{1}{\sqrt{2\tau}}\left(\begin{tabular}{c}
$0$\\
$\sqrt{K+C}\cos{\zeta_{1}}e^{i\vartheta_{1}}$\\
$\sqrt{K-C}\sin{\zeta_{2}}e^{i\vartheta_{2}}$\\
$0$
\end{tabular}\right)
\end{eqnarray}
where all angular functions $\zeta_{1}$, $\zeta_{2}$ and $\theta_{1}$, $\theta_{2}$, $\vartheta_{1}$, $\vartheta_{2}$ have only temporal dependence, that has to be determined by plugging the spinor back into the Dirac equations.
\\
For the gravitational field, borrowing again from \cite{Saha2}, we subtract equation \eqref{3.10b} from equation \eqref{3.10d} obtaining
\begin{equation}\label{3.14.0}
\varphi\tau\frac{d}{dt}\left(\frac{\dot a}{a} - \frac{\dot c}{c}\right) + \varphi\dot\tau\left(\frac{\dot a}{a} - \frac{\dot c}{c}\right) + \dot{\varphi}\tau\left(\frac{\dot a}{a} - \frac{\dot c}{c}\right)=0
\end{equation}
from which we derive
\begin{equation}\label{3.14}
\frac{a}{c}=De^{\left(X\int{\frac{dt}{\varphi\tau}}\right)}
\end{equation}
$D$ and $X$ being suitable constants, which gives the evolution equation for the ratio of the two axes that are not constantly proportional, and therefore it can be considered as the evolution equation for the shape of the universe; also, multiplying \eqref{3.10a} by $3$ and adding the result to the summation of equations \eqref{3.10b}, \eqref{3.10c} and \eqref{3.10d}, we get the final equation for $\tau$
\begin{equation}\label{3.15}
2\frac{\ddot\tau}{\tau} + 3\frac{\ddot\varphi}{\varphi} = 3\frac{\rho}{\varphi} - 3\frac{p}{\varphi} + \frac{3}{2\varphi}m\bar\psi\psi - 5\frac{\dot\tau}{\tau}\frac{\dot\varphi}{\varphi} - \frac{6}{\varphi^2}V\/(\varphi)
\end{equation}
which can be considered as the evolution equation of the volume of the universe. Here, it is worth noticing that equation \eqref{3.10a} plays the role of a constraint on the initial data: thus for consistency we have to check that, if satisfied initially, this constraint is preserved in time. To see this point, we first observe that the Einstein-like equations \eqref{2.15}, and thus also \eqref{3.10}, can be written in the equivalent form
\begin{equation}\label{3.14.1}
\tilde{R}_{ij}= \tilde{T}_{ij} -\frac{1}{2}\tilde{T}g_{ij}
\end{equation}
where
\begin{equation}\label{3.14.2}
\begin{split}
\tilde{T}_{ij}:= \frac{1}{\varphi}\Sigma^F_{ij} + \frac{1}{\varphi}\tilde{\Sigma}^D_{ij}
+ \frac{1}{\varphi^2}\left( - \frac{3}{2}\de\varphi/de{x^i}\de\varphi/de{x^j} + \varphi\tilde{\nabla}_{j}\de\varphi/de{x^i} + 
\frac{3}{4}\de\varphi/de{x^h}\de\varphi/de{x^k}g^{hk}g_{ij} \right. \\
\left. - \varphi\tilde{\nabla}^h\de\varphi/de{x^h}g_{ij} - V\/(\varphi)g_{ij} \right) + \frac{3}{64\varphi^2}(\bar{\psi}\gamma_5\gamma^\tau\psi)(\bar{\psi}\gamma_5\gamma_\tau\psi)g_{ij}
\end{split}
\end{equation}
denotes the effective stress-energy tensor appearing on the right hand side of equations \eqref{2.15}, while $\tilde T$ is its trace. It is then a straightforward matter to verify that equations \eqref{3.14.0} and \eqref{3.15} can be equivalently obtained by suitably combining the space-space equations of the set \eqref{3.14.1}; more in detail, equation \eqref{3.14.0} is obtained by subtracting each to other the two distinct space-space equations of \eqref{3.14.1}, while equation \eqref{3.15} is obtained adding together all the space-space equations of \eqref{3.14.1}. As a consequence, we have that solving equations \eqref{3.14.0} and \eqref{3.15} amounts to solve all the space-space equations of the set \eqref{3.14.1}. In addition to this, as it is proved in Section 5, the conservation laws \eqref{2.2} automatically imply the vanishing of the four-divergence of the Einstein-like equations \eqref{2.15}. The two just mentioned facts allow to apply to the present case a result by Bruhat (see \cite{yvonne4}, Theorem 4.1, pag. 150) which ensures that the constraint \eqref{3.10a} is actually satisfied for all time.

\paragraph{Examples: $f(R)\equiv R$ and $f(R)\equiv R+\delta\alpha R^{2}$ models.} In the following, we will consider the situation given by $f(R)\equiv R+\delta\alpha R^{2}$ as a correction in $\delta\alpha$ with respect to the $f(R)\equiv R$ case, and we are going to take solutions for the Einstein-like field equations that are corrections in $\delta\alpha$ of the exact solutions of the Einstein field equations: thus the first step is to find the exact solutions of the $f(R)\equiv R$ case.
\\
To this extent, consider for simplicity the simplest case without fluid. Then, we choose for the Dirac field the solution in the form given as in \eqref{restrictedspinor2} by
\begin{eqnarray}
\label{spinorsolution}
&\psi=\frac{1}{\sqrt{2\tau}}\left(\begin{tabular}{c}
$0$\\
$\sqrt{K+C}e^{i\left(-mt-\frac{3C}{16}\int{\frac{dt}{\tau}}\right)}$\\
$\sqrt{K-C}e^{-i\left(-mt-\frac{3C}{16}\int{\frac{dt}{\tau}}\right)}$\\
$0$
\end{tabular}\right)
\end{eqnarray}
with constraints $\bar\psi\psi=\frac{C}{\tau}$, $\psi^{\dagger}\psi=\frac{K}{\tau}$ and $\psi^{\dagger}\gamma^5\psi=0$, $i\bar\psi\gamma^5\psi=0$, and with the solution defined up to a global unitary phase which is then irrelevant: this solution albeit simple is nevertheless not trivial, and it will actually lead to a correspondingly simple form of the gravitational evolution equations for the shape and volume of the universe as 
\begin{equation}
\label{3.18}
\frac{a}{c}=De^{\left(X\int{\frac{1}{\tau}dt}\right)}
\end{equation}
and also
\begin{equation}
\label{3.19}
\ddot\tau-\frac{3mC}{4}=0
\end{equation}
in terms of the constants $m$, $C$, $X$, $D$ alone. As it is clear, it is possible to integrate the equation \eqref{3.19} to give the evolution of the volume of the universe as
\begin{equation}
\label{3.20}
\tau=\frac{3mC}{8}(b+2\beta t +t^{2})
\end{equation}
where $b$ is the integration constant that contains the information about the initial volume of the universe, which has to be positive thus giving rise to a initial state without singularity, and $\beta$ is the integration constant that contains the information about the initial velocity of expansion of the universe; then by recalling that $\tau=a^{2}c$ we can use equation \eqref{3.18} to obtain the evolution of the shape of the universe as in the following
\begin{equation}
\label{3.21}
\frac{a}{c}=D\left(\frac{t+\beta+\sqrt{\beta^{2}-b}}{t+\beta-\sqrt{\beta^{2}-b}}\right)^{-\frac{8X}{6mC\sqrt{\beta^{2}-b}}}
\end{equation}
in which we see that for $t$ that tends to infinity the two factors tend to become constantly proportional, giving isotropization: by setting then $D=1$ and choosing for simplicity $8X=-3mC\sqrt{\beta^{2}-b}$ the evolution of the single factors can be given as in the following
\begin{subequations}
\label{3.22}
\begin{equation}
\label{3.22a}
a=\sqrt[3]{\frac{3mC}{8}}\left(t+\beta-\sqrt{\beta^{2}-b}\right)^{-\frac{1}{3}}
\left(b+2\beta t +t^{2}\right)^{\frac{1}{2}}
\end{equation}
\begin{equation}
\label{3.22c}
c=\sqrt[3]{\frac{3mC}{8}}\left(t+\beta-\sqrt{\beta^{2}-b}\right)^{\frac{2}{3}}
\end{equation}
\end{subequations} 
with constraint given by $3m^{2}(\beta^{2}-b)=1$ for consistency with the fact that the time-time gravitational field equation is not actually a field equation but a constraint. Finally the spinor evolves as
\begin{eqnarray}
&\psi=\frac{2}{\sqrt{3m(b+2\beta t +t^{2})}}\left(\begin{tabular}{c}
$0$\\
$\sqrt{\left(\frac{K}{C}+1\right)}e^{i\left(-mt
+\sqrt{\frac{3}{16}}\ln{\left(\frac{m\sqrt{3}(t+\beta)+1}{m\sqrt{3}(t+\beta)-1}\right)}\right)}$\\
$\sqrt{\left(\frac{K}{C}-1\right)}e^{-i\left(-mt
+\sqrt{\frac{3}{16}}\ln{\left(\frac{m\sqrt{3}(t+\beta)+1}{m\sqrt{3}(t+\beta)-1}\right)}\right)}$\\
$0$
\end{tabular}\right)
\end{eqnarray}
with the spinor bilinears diluting as the inverse of the volume, correspondingly to the fact that they represent densities. By reading this solution we have that: at the first instant of the evolution the universe has no singularities as it occupies a finite volume and it is shaped as a disc whose shorter axis is along the third coordinate described by the $c$ factor, and it is filled with spinor fields with torsional interactions responsible for the anisotropy of the universe; as the time goes by the spinor field dilutes down, allowing isotropization of the universe without relenting its expansion; eventually for very late epochs the scale factors have the $t^{\frac{2}{3}}$ scaling we know for the standard FLRW cosmic evolution.
\\
Now that we have the exact solutions of the $f(R)\equiv R$ case, we can employ them as the basis upon which to build the correction in $\delta\alpha$ for the more general solutions of the $f(R)\equiv R+\delta\alpha R^{2}$.
\\
The form of the Dirac field will be unchanged: its form is going to give rise to a correspondingly simple form of the $\varphi$ function given by 
\begin{equation}
\varphi\equiv 1-\delta\alpha\frac{ m C}{\tau}
\end{equation}
with potential
\begin{equation}
V(\varphi)\equiv \delta\alpha \frac{m^{2} C^{2}}{8\tau^{2}}
\end{equation}
and the gravitational evolution equation for the volume of the universe is
\begin{equation}
\left(\ddot\tau-\frac{3mC}{4}\right)
-\delta\alpha\frac{ mC}{2}\left(\frac{\dot\tau^{2}}{\tau^{2}}-\frac{\ddot\tau}{\tau}
-\frac{3mC}{4\tau}\right)=0
\end{equation}
up to higher-order powers of $\delta\alpha$. As before, it is possible to integrate it to give the evolution of the volume of the universe as
\begin{eqnarray}
&\tau=\frac{3mC}{8}(b\!+\!2\beta t\!+\!t^{2})
+\delta\alpha \frac{mC}{2}\left(\varsigma\!\left(\xi\!+\!t\right)\!
+\!m\sqrt{3}\left(\beta\!+\!t\right)
\ln{\left(\frac{m\sqrt{3}(t+\beta)+1}{m\sqrt{3}(t+\beta)-1}\right)}\right)
\end{eqnarray}
where $\varsigma$ and $\xi$ are two new integration constants for which the initial volume of the universe is now different, but of course there is still no singularity: again these two new integration constants are linked by a relationship coming from the fact that the time-time gravitational field equation is a constraint. We remark that at later cosmological times in the evolution of the universe we still have isotropization, but more in general we have that this solution tends to be approximated to the same we would have had in the Einstein theory in what could be called Einsteinization.
%%%%%%%%%%%%%%%%%%%%%%%%%%%%%%%%%%%%%%%%%%%%%%%%%%%%%%%%%%%%%%%%%%%%%%%%%%%%%%%%%%%%%%%%%%%%%%%%%%
\section{From the Jordan to the Einstein frame}
In this section we shall show that, by passing from the Jordan ($g_{ij}$) to the Einstein frame ($\bar{g}_{ij}$) through a conformal transformation
of the kind $\bar{g}_{ij}=\varphi\/g_{ij}$ (if $\varphi > 0$) or $\bar{g}_{ij}=-\varphi\/g_{ij}$ (if $\varphi< 0$), it is possible to quadratically integrate the Einstein-like equations \eqref{3.10} also in the general case $f(R)\not = R$. 
\\
To start with, supposing for simplicity $\varphi >0$, we note that the relations between the components of the tetrad fields associated with the metric tensors $\bar{g}_{ij}$ and $g_{ij}$ are expressed as
\begin{equation}\label{4.1}
\bar{e}^\mu_i =\sqrt{\varphi}e^\mu_i, \qquad \bar{e}^i_\mu = \frac{1}{\sqrt{\varphi}}e^i_\mu
\end{equation}
while the relation between the Levi--Civita connections induced by the two different frames is given by  
\begin{equation}\label{4.2}
\bar{\Gamma}_{ij}^{\;\;\;h}= \tilde{\Gamma}_{ij}^{\;\;\;h} +
\frac{1}{2\varphi}\de\varphi/de{x^j}\delta^h_i -
\frac{1}{2\varphi}\de\varphi/de{x^p}g^{ph}g_{ij} +
\frac{1}{2\varphi}\de\varphi/de{x^i}\delta^h_j\,.
\end{equation}
Moreover, we have also the identities
\begin{equation}\label{4.3}
\Gamma^h = e^h_\mu\/\gamma^\mu = \sqrt{\varphi}\bar{e}^h_\mu\/\gamma^\mu = \sqrt{\varphi}\bar{\Gamma}^h, \qquad \Gamma_h = \frac{1}{\sqrt{\varphi}}\bar{\Gamma}_h
\end{equation}
between the matrices $\Gamma$ and $\bar\Gamma$ associated with the tetrads $e$ and $\bar e$ respectively. 
\\
Our aim is now to express the Dirac and Einstein-like equations in terms of the conformally transformed metric $\bar{g}_{ij}$ and its tetrad field $\bar{e}^\mu_i$. The covariant derivatives of the spinor fields are defined as in \eqref{3.7}; in order to correctly evaluate them in the Einstein frame, we need to express the coefficients $\tilde{\Omega}_i$ using the metric $\bar{g}_{ij}$ and its related quantities. With respect to this point, taking equations
\eqref{2.3}, \eqref{4.1}, \eqref{4.2} and \eqref{4.3} as well as the identity
\begin{equation}\label{4.4}
- e^j_\mu\partial_i\/e^\mu_k = - \bar{e}^j_\mu\partial_i\/\bar{e}^\mu_k + \frac{1}{2\varphi}\partial_i\varphi\delta^j_k
\end{equation}
into account, we get the representation
\begin{equation}\label{4.5}
\tilde{\Omega}_i = - \frac{1}{4}\bar{g}_{jh}\left(\bar{\Gamma}_{ik}^{\;\;\;j} - \bar{e}^j_\mu\partial_i\/\bar{e}^\mu_k - \frac{1}{2\varphi}\partial_k\varphi\delta^j_i + \frac{1}{2\varphi}\partial_p\varphi\/\bar{g}^{pj}\bar{g}_{ik} \right)\bar{\Gamma}^h\bar{\Gamma}^k
\end{equation}
Now, supposing that the metric $\bar{g}_{ij}$ is of the Bianchi type-I form (the metric $g_{ij}$ is then of the Bianchi type-I only up to a time reparametrization $d\bar{t}=\frac{1}{\sqrt{\varphi}}dt$) 
\begin{equation}\label{4.6}
ds^2 = dt^2 - \bar{a}^2(t)\,dx^2 - \bar{b}^2\,(t)dy^2 - \bar{c}^2(t)\,dz^2
\end{equation}
the quantities \eqref{4.5} assume the explicit expression
\begin{equation}\label{4.7}
\begin{split}
\tilde{\Omega}_0=0,\qquad \tilde{\Omega}_1= \frac{1}{2}\dot{\bar a}\gamma^1\gamma^0 - \frac{\dot{\varphi}}{4\varphi}\bar{a}\gamma^1\gamma^0 := \bar{\Omega}_1 - \frac{\dot{\varphi}}{4\varphi}\bar{a}\gamma^1\gamma^0, \\
\tilde{\Omega}_2= \frac{1}{2}\dot{\bar b}\gamma^2\gamma^0 - \frac{\dot{\varphi}}{4\varphi}\bar{b}\gamma^2\gamma^0 := \bar{\Omega}_2 - \frac{\dot{\varphi}}{4\varphi}\bar{b}\gamma^2\gamma^0, \\
\tilde{\Omega}_3= \frac{1}{2}\dot{\bar c}\gamma^3\gamma^0 - \frac{\dot{\varphi}}{4\varphi}\bar{c}\gamma^3\gamma^0 := \bar{\Omega}_3 - \frac{\dot{\varphi}}{4\varphi}\bar{c}\gamma^3\gamma^0
\end{split}
\end{equation}
Inserting equations \eqref{4.3} and \eqref{4.7} into equations \eqref{2.16}, we obtain the following representation of the Dirac equations 
\begin{subequations}\label{4.8}
\begin{equation}\label{4.8a}
\dot\psi + \frac{\dot{\bar\tau}}{2\bar\tau}\psi - \frac{3\dot\varphi}{4\varphi}\psi + \frac{im}{\sqrt{\varphi}}\gamma^0\psi + \frac{3i}{16\varphi\sqrt{\varphi}}\left[ (\bar\psi\psi)\gamma^0 + i(i\bar\psi\gamma^5\psi)\gamma^0\gamma^5 \right]\psi =0 
\end{equation} 
\begin{equation}\label{4.8b}
\dot{\bar\psi} + \frac{\dot{\bar\tau}}{2\bar\tau}\bar\psi - \frac{3\dot\varphi}{4\varphi}\bar\psi - \frac{im}{\sqrt{\varphi}}\bar{\psi}\gamma^0 - \frac{3i}{16\varphi\sqrt{\varphi}}\bar{\psi}\/\left[ (\bar\psi\psi)\gamma^0 + i(i\bar\psi\gamma^5\psi)\gamma^5\gamma^0 \right] =0 
\end{equation} 
\end{subequations}
where we have defined $\bar{\tau}:=\bar{a}\bar{b}\bar{c}$. Proceeding as in the Jordan frame, from equations \eqref{4.8} we derive the differential equations 
\begin{subequations}\label{4.9}
\begin{equation}\label{4.9a}
\frac{d}{dt}[\bar{\tau}\Theta(\bar\psi\psi)] + \frac{3}{8\varphi}\bar{\tau}(i\bar\psi\gamma^5\psi)(\bar\psi\gamma^5\gamma^0\psi) =0
\end{equation}
\begin{equation}\label{4.9b}
\frac{d}{dt}[\bar{\tau}\Theta(i\bar\psi\gamma^5\psi)] -\frac{2m\bar{\tau}\Theta}{\sqrt{\varphi}}\/(\bar\psi\gamma^5\gamma^0\psi) - \frac{3}{8\varphi}\bar{\tau}(\bar\psi\psi)(\bar\psi\gamma^5\gamma^0\psi) =0
\end{equation}
\begin{equation}\label{4.9c}
\frac{d}{dt}[\bar{\tau}\Theta(\bar\psi\gamma^5\gamma^0\psi)] + \frac{2m\bar{\tau}\Theta}{\sqrt{\varphi}}\/(i\bar\psi\gamma^5\psi) =0
\end{equation}
\end{subequations}
where $\Theta:= \varphi^{-\frac{3}{2}}$. Finally, equations \eqref{4.9} yield the relation (analogous of equation \eqref{3.13bis})
\begin{equation}\label{4.10}
(\bar\psi\psi)^2 + (i\bar\psi\gamma^5\psi)^2 + (\bar\psi\gamma^5\gamma^0\psi)^2 = \frac{C^2}{\bar{\tau}^2\Theta^2}
\end{equation}
$C$ being a constant. 
\\
Now, we deal with the Einstein-like equations \eqref{2.15}. As it has been shown in \cite{CCSV1,CV4,CV1,CV2,CV3}, equations \eqref{2.15} can be expressed in terms of the metric $\bar{g}_{ij}$, assuming the form 
\begin{equation}\label{4.11}
\bar{R}_{ij} -\frac{1}{2}\bar{R}\bar{g}_{ij}= \frac{1}{\varphi}\tilde{\Sigma}^D_{ij} + \frac{1}{\varphi}\Sigma^F_{ij} - \frac{1}{\varphi^3}V\/(\varphi)\bar{g}_{ij} + \frac{3}{64\varphi^3}(\bar{\psi}\gamma_5\gamma^\tau\psi)(\bar{\psi}\gamma_5\gamma_\tau\psi)\bar{g}_{ij}
\end{equation}
where $\bar{R}_{ij}$ and $\bar{R}$ denote the Ricci tensor and the Ricci scalar curvature induced by the metric $\bar{g}_{ij}$. Of course, also the tensors $\tilde{\Sigma}^D_{ij}$ and $\Sigma^F_{ij}$ have to be represented making use of the metric $\bar{g}_{ij}$ and the associated tetrad $\bar{e}^\mu_i$. In this regard, we have
\begin{equation}\label{4.12}
\Sigma^F_{ij}:= \frac{1}{\varphi}(\rho +p)\/\bar{U}_i\bar{U}_j -\frac{p}{\varphi}\bar{g}_{ij}
\end{equation}
with $\bar{U}_i\bar{U}_j\/\bar{g}^{ij}=1$. Concerning the tensor $\tilde{\Sigma}^D_{ij}$, from a direct calculation we obtain the identities
\begin{equation}\label{4.13}
\tilde{\Sigma}^D_{00}= \frac{1}{2\varphi}\left[ m\bar\psi\psi - \frac{3}{16\varphi}(\bar{\psi}\gamma_5\gamma^\nu\psi)(\bar{\psi}\gamma_5\gamma_\nu\psi)\right], \qquad \tilde{\Sigma}^D_{AA}=0 \quad A=1,2,3
\end{equation}
while, due to equations \eqref{4.11}, the vanishing of the non diagonal part of $\tilde{\Sigma}^D_{ij}$ gives rise to constraints analogous to those existing in the Jordan frame, namely
\begin{subequations}\label{4.14}
\begin{equation}\label{4.14a}
\tilde{\Sigma}^D_{12}=0\quad \Rightarrow \quad \bar{a}\dot{\bar{b}} - \bar{b}\dot{\bar{a}}=0 \quad \cup \quad \bar\psi\gamma^5\gamma^3\psi =0
\end{equation}
\begin{equation}\label{4.14b}
\tilde{\Sigma}^D_{23}=0\quad \Rightarrow \quad \bar{c}\dot{\bar{b}} - \bar{b}\dot{\bar{c}}=0 \quad \cup \quad \bar\psi\gamma^5\gamma^1\psi =0
\end{equation}
\begin{equation}\label{4.14c}
\tilde{\Sigma}^D_{13}=0\quad \Rightarrow \quad \bar{a}\dot{\bar{c}} - \bar{c}\dot{\bar{a}}=0 \quad \cup \quad \bar\psi\gamma^5\gamma^2\psi =0
\end{equation}
\end{subequations}
As in the Jordan frame, it is also seen that $\tilde{\Sigma}^D_{0A}=0$ ($A=1,2,3$) are trivial identities, yielding no restrictions.
\\
Collecting all the obtained results, we conclude that in the Einstein frame \eqref{4.6} the Einstein-like equations \eqref{2.15} are expressed as
\begin{subequations}\label{4.15}
\begin{equation}\label{4.15a}
\frac{\dot{\bar a}}{\bar a}\frac{\dot{\bar b}}{\bar b} + \frac{\dot{\bar b}}{\bar b}\frac{\dot{\bar c}}{\bar c} + \frac{\dot{\bar a}}{\bar a}\frac{\dot{\bar c}}{\bar c} = \frac{\rho}{\varphi^2} +
\frac{1}{2\varphi^2}m\bar\psi\psi - \frac{3}{64\varphi^3}(\bar{\psi}\gamma_5\gamma^\tau\psi)(\bar{\psi}\gamma_5\gamma_\tau\psi) - 
\frac{1}{\varphi^3}V(\varphi) 
\end{equation}
\begin{equation}\label{4.15b}
\frac{\ddot{\bar b}}{\bar b} + \frac{\ddot{\bar c}}{\bar c} + \frac{\dot{\bar b}}{\bar b}\frac{\dot{\bar c}}{\bar c} = - \frac{p}{\varphi^2} -
\frac{1}{\varphi^3}V(\varphi) + \frac{3}{64\varphi^3}(\bar{\psi}\gamma_5\gamma^\tau\psi)(\bar{\psi}\gamma_5\gamma_\tau\psi)
\end{equation}
\begin{equation}\label{4.15c}
\frac{\ddot{\bar a}}{\bar a} + \frac{\ddot{\bar c}}{\bar c} + \frac{\dot{\bar a}}{\bar a}\frac{\dot{\bar c}}{\bar c} = - \frac{p}{\varphi^2} - 
\frac{1}{\varphi^3}V(\varphi) + \frac{3}{64\varphi^3}(\bar{\psi}\gamma_5\gamma^\tau\psi)(\bar{\psi}\gamma_5\gamma_\tau\psi)
\end{equation}
\begin{equation}\label{4.15d}
\frac{\ddot{\bar a}}{\bar a} + \frac{\ddot{\bar b}}{\bar b} + \frac{\dot{\bar a}}{\bar a}\frac{\dot{\bar b}}{\bar b} = - \frac{p}{\varphi^2} - 
\frac{1}{\varphi^3}V(\varphi) + \frac{3}{64\varphi^3}(\bar{\psi}\gamma_5\gamma^\tau\psi)(\bar{\psi}\gamma_5\gamma_\tau\psi)
\end{equation}
\end{subequations}
together with the constraints \eqref{4.14}. Again, the only significant case is when two components among $\bar{a}(t)$, $\bar{b}(t)$ and $\bar{c}(t)$ are equal. For instance, we suppose $\bar{a}(t)=\bar{b}(t)$ and thus $\bar\psi\gamma^5\gamma^1\psi = \bar\psi\gamma^5\gamma^2\psi =0$. We can now proceed as in the Jordan frame (see section 3), so obtaining from equations \eqref{4.15} the relations 
\begin{equation}\label{4.16}
\frac{\bar a}{\bar c} = D_1\exp\left(X_1\int{\frac{1}{\bar\tau}dt}\right)
\end{equation}
\begin{equation}\label{4.17}
2\frac{\ddot{\bar\tau}}{\bar\tau} = 3\frac{\rho}{\varphi^2} - 3\frac{p}{\varphi^2} + \frac{3}{2\varphi^2}m\bar\psi\psi - \frac{6}{\varphi^3}V\/(\varphi)
\end{equation}
The last equations we have to express in terms of the metric $\bar{g}_{ij}$ are those concerning the conservation law for the perfect fluid
\begin{equation}
\label{4.18}
\tilde{\nabla}_i\Sigma_F^{ij} =0
\end{equation}
with
\begin{equation}
\label{4.18bis}
\Sigma_F^{ij}= \varphi(\rho +p)\bar{U}^i\bar{U}^j - \varphi\/p\/\bar{g}^{ij}
\end{equation}
Taking equations \eqref{4.2} and \eqref{4.6} into account, it is easily seen that equations \eqref{4.18} and \eqref{4.18bis} reduce to
\begin{equation}\label{4.19}
\frac{\dot\rho}{\rho} + (1+\lambda)\frac{\dot{\bar\tau}}{\bar\tau} - \frac{3}{2}\frac{\dot\varphi}{\varphi}(1+\lambda)=0
\end{equation}
Setting $\Theta:= \varphi^{-\frac{3}{2}}$, one has $- \frac{3}{2}\frac{\dot\varphi}{\varphi}=\frac{\dot\Theta}{\Theta}$ 
and we can integrate equation \eqref{4.19} as
\begin{equation}\label{4.20}
\rho =\rho_0\/(\bar{\tau}\Theta)^{-(1+\lambda)} \qquad \rho_0= {\rm constant}
\end{equation}
At this point, we notice that if we could express the quantities $\rho$, $\bar\psi\psi$ e $\varphi$ (or, equivalently, $\Theta=\varphi^{-\frac{3}{2}}$) as functions of $\bar\tau$ and insert the so obtained results into equation \eqref{4.17}, we would get a final equation for $\bar\tau$ of the kind 
\begin{equation}\label{4.21}
\ddot{\bar\tau}= f(\bar\tau)
\end{equation}
which is quadratically integrable. On the other hand, there exists a special though not too restrictive case for which the above mentioned conditions are satisfied. Indeed, setting $i\bar\psi\gamma^5\psi=\bar\psi\gamma^5\gamma^0\psi=0$ in \eqref{4.10}, we obtain
\begin{equation}\label{4.22}
\bar\psi\psi = \frac{C}{\bar{\tau}\Theta}
\end{equation}
equation \eqref{4.22} together with \eqref{4.20} and $\Theta^{-\frac{2}{3}}=\varphi=f'(R(\bar\psi\psi,\rho))$
constitute a set of three relations involving the four variables $\bar\tau$, $\bar\psi\psi$, $\rho$ e $\Theta$. In principle, it is then possible to express the last three variables as suitable functions of $\bar\tau$ alone, thus obtaining from \eqref{4.17} a final quadratically integrable equation of the kind \eqref{4.21}.
\\
As a last remark, we note that equation \eqref{4.15a} has the nature of a constraint on the initial data. As made in the Jordan frame, we have again to check that such a constraint is satisfied for all time after the initial instant. In connection with this, in Section 5 it is proved that the conservation laws \eqref{2.2} ensure that the four-divergence (with respect to the Levi--Civita connection associated with the conformal metric $\bar{g}_{ij}$) of the effective stress-energy tensor
\begin{equation}\label{4.23}
T_{ij}:= \frac{1}{\varphi}\tilde{\Sigma}^D_{ij} + \frac{1}{\varphi}\Sigma^F_{ij} - \frac{1}{\varphi^3}V\/(\varphi)\bar{g}_{ij} + \frac{3}{64\varphi^3}(\bar{\psi}\gamma_5\gamma^\tau\psi)(\bar{\psi}\gamma_5\gamma_\tau\psi)\bar{g}_{ij}
\end{equation}
vanishes. Repeating the arguments developed in section 3, we can apply again the Bruhat's result \cite{yvonne4} and thus verify that the Hamiltonian constraint \eqref{4.15a} is preserved in time. 
\\
We now illustrate two explicit examples where the above procedure actually works:
\paragraph{Example 1: $f(R)\equiv R^{4}$ model.} To begin with, let us consider the model $f(R)=R^4$ coupled with a Dirac field (without perfect fluid for simplicity). In such a case, the corresponding potential \eqref{2.8} assumes the form
\begin{equation}\label{6.1}
V\/(\varphi)= \left(\frac{1}{4}\right)^{\frac{1}{3}}\frac{3}{8}\varphi^{\frac{7}{3}}
\end{equation}
with the scalar field \eqref{2.7} given by
\begin{equation}\label{6.2}
\varphi = 4\left(\frac{m}{4}\bar\psi\psi\right)^{\frac{3}{4}}
\end{equation}
Through equation \eqref{4.22} and the definition of $\Theta=\varphi^{-\frac{3}{2}}$, we get the relation
\begin{equation}
\label{6.3}
\Theta = A\/\left(\frac{C}{\bar\tau}\right)^9
\end{equation}
with
\begin{equation}
\label{6.3bis}
A=\left[4\/\left(\frac{m}{4}\right)^{\frac{3}{4}}\right]^{12}
\end{equation}
from which, using again \eqref{4.22}, we have
\begin{equation}\label{6.4}
\frac{3}{2\varphi^2}m\bar\psi\psi= \frac{3m}{32C^8\/\left(\frac{m}{4AC^8}\right)^{\frac{3}{4}}}\bar{\tau}^{-4}
\end{equation}
and 
\begin{equation}\label{6.5}
-\frac{6}{\varphi^3}V\/(\varphi) = -\frac{9}{4}\left(\frac{1}{4}\right)^{\frac{5}{6}}\left(\frac{m}{4AC^8}\right)^{-\frac{1}{2}}\bar{\tau}^{-4}
\end{equation}
Inserting equations \eqref{6.4} and \eqref{6.5} into \eqref{4.17} and choosing the constant $C$ sufficiently small, we obtain the final differential equation
\begin{equation}\label{6.6}
\ddot{\bar\tau}= B\bar{\tau}^{-3}
\end{equation}
with $B>0$. The latter can be easily integrated as
\begin{equation}\label{6.7}
\bar\tau = \sqrt{D\/(t+E)^2 + \frac{B}{D}}
\end{equation}
where the constants $D>0$ and $E$ have to be determined consistently with equation \eqref{4.15a}. From equations \eqref{6.6} and \eqref{6.7} we have also $\bar{\tau}^2 \geq \frac{B}{D}$. We see that, at least in the Einstein frame, the model $f(R)=R^4$ coupled  with just one Dirac Field gives rise to a universe with no initial singularity but undergoing a slow expansion without isotropization. 

\paragraph{Example 2: $f(R)\equiv R+\alpha R^{2}$ model.} The second example we consider is given by the model $f(R)=R+\alpha\/R^2$ with $\alpha < 0$, still coupled with a Dirac field alone. The associated potential \eqref{2.8} is now of the form
\begin{equation}\label{6.8}
V(\varphi)= \frac{1}{8\alpha}\varphi(\varphi -1)^2
\end{equation}
while the scalar field \eqref{2.7} is 
\begin{equation}\label{6.9}
\varphi=1 - \alpha\/m\/\bar\psi\psi
\end{equation}
From equation \eqref{4.22} and the definition of $\Theta$ we get the algebraic equation of third order (for the unknown $\Theta\bar\tau$)
\begin{equation}\label{6.10}
\left(\bar\tau\/\Theta - \alpha\/mC 
 \right) ^{3} - {\bar\tau}^{3}\/\Theta=0
\end{equation}
linking $\Theta$ to the volume-scale $\bar\tau$. We recall that we  are working under the assumption $\varphi >0$. We are then forced to require $\Theta >0$. In order to satisfy this condition, we have to impose that equation \eqref{6.10} has three real solutions (indeed, if equation \eqref{6.10} had only one real solution, the latter would be negative as it is easily seen solving \eqref{6.10} by a graph). The imposed requirement yields the condition $\bar\tau \geq \frac{3\sqrt{3}}{2}|\alpha|\/mC$ which represents a lower bound for the volume-scale $\bar\tau$. Therefore, again we can avoid initial singularity. Anyway, a solution remaining positive in time is 
\begin{equation}\label{6.11}
\Theta\bar\tau=\alpha\/mC+\frac{2}{3}\sqrt {3}\bar\tau\/\cos\left(\frac{1}{3}\phi \right)
\end{equation}
where
\begin{equation}
\phi = \arccos \left( \frac{3}{2}{\frac{\sqrt{3}\left| \alpha \right|\/mC}{\bar\tau}} \right)
\end{equation} 
At this point, choosing parameters such that $\alpha= - \frac{1}{2}$ and $mC=1$ for simplicity, and inserting equations \eqref{6.10} and \eqref{6.11} into \eqref{4.17}, after some calculation we end up with the final differential equation
\begin{equation}\label{6.12}
\ddot{\bar\tau} =\frac{\sqrt {3}}{4}\cos
 \left( \frac{1}{3}\,\arccos \left( \frac {3\sqrt {3}}{4\bar\tau} \right)  \right) + \frac{3\sqrt {3}}{16}\left( \cos \left( \frac{1}{3}
\arccos \left( \frac{3\sqrt{3}}{4\bar\tau}
 \right)  \right)  \right) ^{-1}
\end{equation}
Equation \eqref{6.12} is quadratically integrable, but explicit calculations are not so easy. We can then proceed to numerically plotting it. For instance, if initial data $\bar\tau(0)=\frac{3\sqrt{3}}{4}$ and $\dot{\bar\tau}(0)= 1$ are assigned, the corresponding behavior of the volume-scale $\bar\tau(t)$ during time is given by the graph in Figure 1
\begin{figure}[htb]
\includegraphics[height=7.5cm]{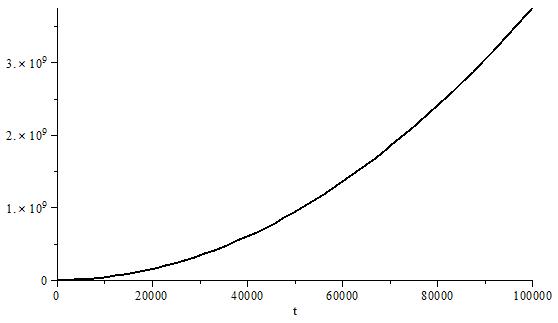}
\caption{volume-scale $\bar\tau$ vs time t}
\label{fig:1}
\end{figure}
showing an (at least initially) accelerated expansion of the volume-scale $\bar\tau$. Moreover, by plotting the function $\frac{\bar\tau}{t}$, it is seen that the volume-scale $\bar\tau$ in infinite of order $n>1$. Due to equation \eqref{4.16}, this fact implies that the universe isotropizes at large $t$. Also, we notice that at large $t$ one has $\varphi \approx 1$ and the Einstein and Jordan frames tend to coincide.
%%%%%%%%%%%%%%%%%%%%%%%%%%%%%%%%%%%%%%%%%%%%%%%%%%%%%%%%%%%%%%%%%%%%%%%%%%%%%%%%%%%%%%%%%%%%%%%%%%%
\section{Conservation laws}
We clarify the relationship between the conservation laws \eqref{2.2} and the Einstein-like equations \eqref{2.15}. More in particular, we show that the validity the conservation laws \eqref{2.2} implies that the four-divergence of the Einstein-like equations \eqref{2.15}, with respect to the Levi--Civita connection, vanishes and vice-versa. Also, we prove that the same result holds in the Einstein-frame, namely the four-divergence of the conformally transformed Einstein-like equations \eqref{4.11}, with respect to the Levi--Civita connection associated with the conformal metric $\bar{g}_{ij}=\varphi\/g_{ij}$, is zero.
\\
To start with, for convenience of the reader, we rewrite equations \eqref{2.2}
\begin{subequations}\label{A.1}
\begin{equation}\label{A.1a}
\nabla_{i}\Sigma^{ij}+T_{i}\Sigma^{ij}-\Sigma_{ih}T^{jih}-\frac{1}{2}S_{hiq}R^{hiqj}=0
\end{equation}
\begin{equation}\label{A.1b}
\nabla_{h}S^{ijh}+T_{h}S^{ijh}+\Sigma^{ij}-\Sigma^{ji}=0
\end{equation}
\end{subequations}
where $\Sigma^{ij}$ denotes the sum of the stress-energy tensors of the Dirac field and the perfect fluid. The curvature tensor can be decomposed in the form
\begin{equation}\label{A.2}
R^h_{\;\;iqj}= \tilde{R}^h_{\;\;iqj} +
\tilde{\nabla}_jK_{qi}^{\;\;\;h} -
\tilde{\nabla}_qK_{ji}^{\;\;\;h} +
K_{ji}^{\;\;\;p}K_{qp}^{\;\;\;h} -
K_{qi}^{\;\;\;p}K_{jp}^{\;\;\;h}
\end{equation}
where the contorsion tensor $K_{ij}^{\;\;\;h}$ is expressed as the sum \cite{CCSV2,FV1}
\begin{equation}\label{A.3}
K_{ij}^{\;\;\;h}= \hat{K}_{ij}^{\;\;\;h} + \hat{S}_{ij}^{\;\;\;h}
\end{equation}
with
\begin{subequations}\label{A.4}
\begin{equation}\label{A.4a}
\hat{S}_{ij}^{\;\;\;h}:=-\frac{1}{2\varphi}\/S_{ij}^{\;\;\;h}
\end{equation}
\begin{equation}\label{A.4b}
\hat{K}_{ij}^{\;\;\;h} := -\hat{T}_j\delta^h_i + \hat{T}_pg^{ph}g_{ij}
\end{equation}
\begin{equation}\label{A.4c}
\hat{T}_j:=\frac{1}{2\varphi}\/\de{\varphi}/de{x^j}
\end{equation}
\end{subequations}
Taking equations \eqref{A.2}, \eqref{A.3} and \eqref{A.4} into account, a direct calculation shows that
\begin{eqnarray}\label{A.5}
&\nonumber
-\frac{1}{2}S_{hiq}R^{hiqj} = - S_h^{\;\;ip}\hat{S}^j_{\;\;ip}\hat{T}^h - \frac{1}{2}\varphi\tilde\nabla^j\/\left(\hat{S}_{qih}\hat{S}^{qih}\right)-\\ &-\varphi\tilde\nabla_i\/\left(\hat{S}^{hqi}\hat{S}^j_{\;\;qh}\right)
+\varphi\tilde\nabla_i\/\left(\hat{S}^{hqi}\right)\/\hat{S}^j_{\;\;qh} 
\end{eqnarray}
Moreover, making use of equations \eqref{2.12} and \eqref{2.13}, it is easily seen that
\begin{equation}\label{A.6}
\nabla_{i}\Sigma^{ij}+T_{i}\Sigma^{ij}-\Sigma_{ih}T^{jih} = \tilde\nabla_{i}\Sigma^{(ij)} + \tilde\nabla_{i}\Sigma^{[ij]} - K_{ih}^{\;\;\;j}\Sigma^{[ih]} - T^j_{\;\;ih}\Sigma^{[ih]}
\end{equation}
In \cite{FV1} it is proved that $\Sigma^{(ij)} = \tilde{\Sigma}^{(ij)} - \varphi\hat{S}^{hip}\hat{S}^j_{\;\;ph}$ and also that the conservation laws \eqref{A.1b} amount to the identities $\frac{1}{\varphi}\Sigma^{[ij]} + \tilde\nabla_h\hat{S}^{jih}$. From this and equations \eqref{A.3} and \eqref{A.4} one gets
\begin{subequations}\label{A.7}
\begin{equation}\label{A.7a}
\tilde\nabla_i\Sigma^{(ij)} = \tilde\nabla_i\tilde{\Sigma}^{(ij)} - \tilde\nabla_i\/\left(\varphi\hat{S}^{hip}\hat{S}^{j}_{\;\;ph}\right)
\end{equation}
\begin{equation}\label{A.7b}
\tilde\nabla_i\Sigma^{[ih]} = - \varphi_i\tilde\nabla_h\hat{S}^{jih}
\end{equation}
\begin{equation}\label{A.7c}
- K_{ih}^{\;\;\;j}\Sigma^{[ih]} = - \varphi\/\hat{T}_h\tilde\nabla_q\hat{S}^{hjq} + \varphi\hat{S}_{ih}^{\;\;\;j}\tilde\nabla_q\hat{S}^{hiq}
\end{equation}
\begin{equation}\label{A.7d}
- T^j_{\;\;ih}\Sigma^{[ih]} = \varphi\/\hat{T}_i\tilde\nabla_q\hat{S}^{jiq} - 2\varphi\hat{S}_{jih}\tilde\nabla_q\hat{S}^{hiq}
\end{equation}
\end{subequations}
where for simplicity we have defined $\varphi_i := \de\varphi/de{x^i}$. Inserting equations \eqref{A.5}, \eqref{A.6} and \eqref{A.7} into \eqref{A.1a}, the latter reduce to 
\begin{equation}\label{A.8}
\tilde\nabla_i\tilde{\Sigma}^{(ij)} - \frac{1}{2}\varphi\tilde\nabla^j\left(\hat{S}_{hqp}\hat{S}^{hqp}\right) =0
\end{equation} 
At this point, we recall that equations \eqref{2.7} and \eqref{2.8} are equivalent to the relation 
\begin{equation}\label{A.9}
\Sigma -\frac{6}{\varphi}V(\varphi) + 2V'(\varphi)=0
\end{equation}
(see \cite{CCSV1} for the proof). After that, taking the trace of equation \eqref{2.15} as well as the identity $-\frac{1}{2}\hat{S}_{hqp}\hat{S}^{hqp} = \frac{3}{64\varphi^2}(\bar{\psi}\gamma_5\gamma^\tau\psi)(\bar{\psi}\gamma_5\gamma_\tau\psi)$ \cite{FV1} into account, we get 
\begin{equation}\label{A.10}
\tilde\Sigma= -\varphi\tilde{R} - \frac{3}{2}\frac{1}{\varphi}\varphi_i\varphi^i + 3\tilde{\nabla}_i\varphi^i + \frac{4}{\varphi}V(\varphi) + 2\varphi\hat{S}_{hqp}\hat{S}^{hqp}
\end{equation}
Substituting the identity $\Sigma = \tilde\Sigma - \varphi\hat{S}_{hqp}\hat{S}^{hqp}$ \cite{FV1} and equation \eqref{A.10} in equation \eqref{A.9}, we obtain
\begin{equation}\label{A.11}
\tilde{R} + \frac{3}{2}\frac{1}{\varphi^2}\varphi_i\varphi^i - \frac{3}{\varphi}\tilde{\nabla}_i\varphi^i + 
\frac{2}{\varphi^2}V(\varphi) - \frac{2}{\varphi}V'(\varphi) - \hat{S}_{hqp}\hat{S}^{hqp}=0
\end{equation}
We rewrite equation \eqref{2.15} in the form
\begin{equation}\label{A.12}
\begin{split}
\varphi\tilde{R}_{ij} -\frac{\varphi}{2}\tilde{R}g_{ij}= \tilde\Sigma_{(ij)} + \frac{1}{\varphi}\left( - \frac{3}{2}\varphi_i\varphi_j + \varphi\tilde{\nabla}_{j}\varphi_i + \frac{3}{4}\varphi_h\varphi^h\/g_{ij} + \right. \\
\left. - \varphi\tilde{\nabla}^h\varphi_h\/g_{ij} - V\/(\varphi)g_{ij} \right) -\frac{1}{2}\varphi\hat{S}_{hqp}\hat{S}^{hqp}g_{ij}
\end{split}
\end{equation}
The covariant divergence of \eqref{A.12} yields
\begin{equation}\label{A.13}
\begin{split}
(\tilde\nabla^j\varphi)\tilde{R}_{ij} + \varphi\tilde\nabla^j\tilde{G}_{ij} -\frac{1}{2}\tilde{R}\tilde\nabla_i\varphi = \tilde\nabla^j\tilde\Sigma_{(ij)} + \left(\tilde\nabla^j\tilde{\nabla}_{j}\tilde\nabla_i - \tilde\nabla_i\tilde{\nabla}^j\tilde\nabla_j\right)\varphi +\\
+ \tilde\nabla^j\left[\frac{1}{\varphi}\left(-\frac{3}{2}\varphi_i\varphi_j + \frac{3}{4}\varphi_h\varphi^h\/g_{ij} - V\/(\varphi)g_{ij}\right)\right] - \frac{1}{2}\tilde\nabla_i\left(\varphi\hat{S}_{hqp}\hat{S}^{hqp}\right)
\end{split}
\end{equation}
By definition, the Einstein and the Ricci tensors satisfy $\tilde\nabla^j\tilde{G}_{ij}=0$ and $(\tilde\nabla^j\varphi)\tilde{R}_{ij} = \left(\tilde\nabla^j\tilde{\nabla}_{j}\tilde\nabla_i - \tilde\nabla_i\tilde{\nabla}^j\tilde\nabla_j\right)\varphi$. Then equation \eqref{A.13} reduces to
\begin{eqnarray}\label{A.14}
\nonumber
&-\frac{1}{2}\tilde{R}\tilde\nabla_i\varphi = \tilde\nabla^j\tilde\Sigma_{(ij)}+ \tilde\nabla^j\left[\frac{1}{\varphi}\left(-\frac{3}{2}\varphi_i\varphi_j + \frac{3}{4}\varphi_h\varphi^h\/g_{ij} - V\/(\varphi)g_{ij}\right)\right]-\\
&-\frac{1}{2}\tilde\nabla_i\left(\varphi\hat{S}_{hqp}\hat{S}^{hqp}\right)
\end{eqnarray}
Finally, making use of equation \eqref{A.10} it is easily seen that
\begin{equation}\label{A.15}
-\frac{1}{2}\tilde{R}\tilde\nabla_i\varphi = \tilde\nabla^j\left[\frac{1}{\varphi}\left(-\frac{3}{2}\varphi_i\varphi_j + \frac{3}{4}\varphi_h\varphi^h\/g_{ij} - V\/(\varphi)g_{ij}\right)\right] - \frac{1}{2}\varphi_i\hat{S}_{hqp}\hat{S}^{hqp}
\end{equation}
and then equations \eqref{A.13} amount to the equations $\tilde\nabla^j\tilde{\Sigma}_{(ij)} - \frac{1}{2}\varphi\tilde\nabla_i\left(\hat{S}_{hqp}\hat{S}^{hqp}\right) =0$, clearly identical to \eqref{A.8}. 
\\
Now, we denote by 
\begin{equation}\label{B.2}
T_{ij}= \frac{1}{\varphi}\tilde{\Sigma}_{(ij)} - \frac{1}{\varphi^3}V\/(\varphi)\bar{g}_{ij} - \frac{1}{2\varphi}\hat{S}_{hqp}\hat{S}^{hqp}\bar{g}_{ij}
\end{equation}
the effective stress-energy tensor on the right hand side of equations \eqref{4.11}. Indicating by $\bar\nabla_i$ the covariant derivative associated with the conformal metric $\bar{g}_{ij}=\varphi\/g_{ij}$, we show that the condition $\bar{\nabla}^j\/T_{ij}=0\/$ is equivalent to the conservation laws $\tilde{\nabla}^j\tilde\Sigma_{(ij)}- \frac{1}{2}\varphi\tilde\nabla_i\left(\hat{S}_{hqp}\hat{S}^{hqp}\right)=0$. Recalling equations \eqref{4.2}, we have 
\begin{equation}\label{B.3}
\begin{split}
\bar{\nabla}^j\/T_{ij}= \frac{1}{\varphi}g^{sj}\bar{\nabla}_s\/T_{ij}= \frac{1}{\varphi}g^{sj}\left[\tilde{\nabla}_s\/T_{ij} - \frac{1}{2\varphi}\left(\de{\varphi}/de{x^i}\delta^q_s + \de{\varphi}/de{x^s}\delta^q_i - \de{\varphi}/de{x^u}g^{uq}g_{si}\right)T_{qj} +\right. \\
\left. - \frac{1}{2\varphi}\left(\de{\varphi}/de{x^j}\delta^q_s + \de{\varphi}/de{x^s}\delta^q_j - \de{\varphi}/de{x^u}g^{uq}g_{sj}\right)T_{iq}\right]
\end{split}
\end{equation}
We have separately
\begin{equation}\label{B.4}
\begin{split}
\frac{1}{\varphi}g^{sj}\tilde{\nabla}_s\/T_{ij}=\frac{1}{\varphi}g^{sj}\tilde{\nabla}_s\/\left(\frac{1}{\varphi}\tilde\Sigma_{(ij)} - \frac{1}{\varphi^2}V\/(\varphi)g_{ij} - \frac{1}{2}\hat{S}_{hqp}\hat{S}^{hqp}g_{ij}\right)=\\ \frac{1}{\varphi^2}\tilde{\nabla}^j\tilde\Sigma_{(ij)} - \frac{1}{\varphi^3}\de{\varphi}/de{x^s}\tilde\Sigma_{(i}^{\;\;s)} +
- \frac{1}{\varphi}\tilde\nabla_i\left(\frac{1}{\varphi^2}V\/(\varphi)\right) -\frac{1}{2\varphi}\tilde\nabla_i\left(\hat{S}_{hqp}\hat{S}^{hqp}\right)
\end{split}
\end{equation}
\begin{equation}\label{B.5}
\begin{split}
\frac{1}{\varphi}g^{sj}\frac{1}{2\varphi}\left(\de{\varphi}/de{x^i}\delta^q_s + \de{\varphi}/de{x^s}\delta^q_i - \de{\varphi}/de{x^u}g^{uq}g_{si}\right)T_{qj}=\\
=\frac{1}{\varphi}g^{sj}\frac{1}{2\varphi}\left(\de{\varphi}/de{x^i}\delta^q_s + \de{\varphi}/de{x^s}\delta^q_i - \de{\varphi}/de{x^u}g^{uq}g_{si}\right)\times\\
\times\left(\frac{1}{\varphi}\tilde\Sigma_{(qj)} - \frac{1}{\varphi^2}V\/(\varphi)g_{qj} -\frac{1}{2}\hat{S}_{hqp}\hat{S}^{hqp}g_{qj}\right)=\\
=\frac{1}{2\varphi^3}g^{sj}\/\left(\de{\varphi}/de{x^i}\tilde\Sigma_{(sj)} + \de{\varphi}/de{x^s}\tilde\Sigma_{(ij)} - \de{\varphi}/de{x^u}g_{si}\tilde\Sigma^{(u}_{\;\;j)}\right)+\\
-\frac{1}{2\varphi^4}g^{sj}\/\left(\de{\varphi}/de{x^i}V\/(\varphi)g_{sj} + \de{\varphi}/de{x^s}V\/(\varphi)g_{ij} - \de{\varphi}/de{x^u}V\/(\varphi)\delta^u_jg_{si}\right)=\\
-\frac{1}{4\varphi^2}g^{sj}\left(\de{\varphi}/de{x^i}\hat{S}_{hqp}\hat{S}^{hqp}g_{sj} + \de{\varphi}/de{x^s}\hat{S}_{hqp}\hat{S}^{hqp}g_{ij} - \de{\varphi}/de{x^j}\hat{S}_{hqp}\hat{S}^{hqp}g_{si}\right)=\\
\\=\frac{1}{2\varphi^3}\de{\varphi}/de{x^i}\tilde\Sigma - \frac{2}{\varphi^4}\de{\varphi}/de{x^i}V\/(\varphi) - \frac{1}{\varphi^2}\de{\varphi}/de{x^i}\hat{S}_{hqp}\hat{S}^{hqp}
\end{split}
\end{equation}
and
\begin{equation}\label{B.6}
\begin{split}
\frac{1}{\varphi}g^{sj}\frac{1}{2\varphi}\left(\de{\varphi}/de{x^j}\delta^q_s + \de{\varphi}/de{x^s}\delta^q_j - \de{\varphi}/de{x^u}g^{uq}g_{sj}\right)T_{iq}=\\
=\frac{1}{\varphi}g^{sj}\frac{1}{2\varphi}\left(\de{\varphi}/de{x^j}\delta^q_s + \de{\varphi}/de{x^s}\delta^q_j - \de{\varphi}/de{x^u}g^{uq}g_{sj}\right)\times\\
\times\left(\frac{1}{\varphi}\tilde\Sigma_{(iq)} - \frac{1}{\varphi^2}V\/(\varphi)g_{iq} -\frac{1}{2}\hat{S}_{hqp}\hat{S}^{hqp}g_{iq}\right)=\\
=\frac{1}{2\varphi^3}g^{sj}\/\left(\de{\varphi}/de{x^j}\tilde\Sigma_{(is)} + \de{\varphi}/de{x^s}\tilde\Sigma_{(ij)} - \de{\varphi}/de{x^u}g_{sj}\tilde\Sigma^{(u}_{\;\;i)}\right)+\\
-\frac{1}{2\varphi^4}g^{sj}\/\left(\de{\varphi}/de{x^j}V\/(\varphi)g_{si} + \de{\varphi}/de{x^s}V\/(\varphi)g_{ij} - \de{\varphi}/de{x^u}V\/(\varphi)\delta^u_ig_{sj}\right)=\\
=-\frac{1}{4\varphi^2}g^{sj}\left(\de{\varphi}/de{x^j}\hat{S}_{hqp}\hat{S}^{hqp}g_{is} + \de{\varphi}/de{x^s}\hat{S}_{hqp}\hat{S}^{hqp}g_{ij} - \de{\varphi}/de{x^i}\hat{S}_{hqp}\hat{S}^{hqp}g_{sj}\right)=\\
= -\frac{1}{\varphi^3}\de{\varphi}/de{x^s}\tilde\Sigma^{\;\;s)}_{(i} + \frac{1}{\varphi^4}\de{\varphi}/de{x^i}V\/(\varphi) +\frac{1}{2\varphi^2}\de{\varphi}/de{x^i}\hat{S}_{hqp}\hat{S}^{hqp}
\end{split}
\end{equation}
Collecting equations \eqref{B.4}, \eqref{B.5} and \eqref{B.6} and recalling that $\Sigma=\tilde\Sigma - \varphi\hat{S}_{hqp}\hat{S}^{hqp}$, we have then
\begin{equation}\label{B.7}
\begin{split}
\bar{\nabla}^j\/T_{ij}=\frac{1}{\varphi^2}\left(\tilde{\nabla}^j\tilde\Sigma_{(ij)} -\frac{\varphi}{2}\tilde\nabla_i\hat{S}_{hqp}\hat{S}^{hqp}\right)
+ \frac{\varphi_i}{\varphi^3}\left[-\frac{1}{2}\Sigma + \frac{3}{\varphi}V(\varphi) - V'(\varphi)\right]= \\
= \frac{1}{\varphi^2}\left(\tilde{\nabla}^j\tilde\Sigma_{(ij)} -\frac{\varphi}{2}\tilde\nabla_i\hat{S}_{hqp}\hat{S}^{hqp}\right)
\end{split}
\end{equation}
because the identity $-\frac{1}{2}\Sigma + \frac{3}{\varphi}V(\varphi) - V'(\varphi)=0$ holds identically, being equivalent to the definition $\varphi=f'(F(\Sigma))$ \cite{CCSV1}.
%%%%%%%%%%%%%%%%%%%%%%%%%%%%%%%%%%%%%%%%%%%%%%%%%%%%%%%%%%%%%%%%%%%%%%%%%%%%%%%%%%%%%%%%%%%%%%%%%%%
\section{Conclusion}
In the present paper we started from gravitational theories of the $f(R)$ type in which the Ricci scalar $R$ contains torsion as well as the metric field, considering them in the case in which the spacetime is filled with Dirac matter fields; we have employed these models to study anisotropic cosmological models BI: with respect to Einstein gravity, additional gravitational effects arise as a consequence of the non-linearity of the $f(R)$ function and the natural torsion-spin coupling, known to induce centrifugal barriers \cite{s-s}. These non-linear and repulsive centrifugal effects may enforce one another, especially in the case anisotropies are considered; the relationship between torsion and anisotropies of the spacetime has been widely investigated in the literature, and recent accounts are for example those in \cite{m-b,s-r}).
\\
An important issue that must be highlighted is the fact that, despite the anisotropic background, the Einstein tensor is diagonal, while, because of the intrinsic features of the spinor field, the energy tensor is not diagonal: in this circumstance the non-diagonal part of the gravitational field equations results into the constraints \eqref{3.11} characterizing the structure of the spacetime or the helicity of the spinor field or both; in our understanding, the only physically meaningful situation is the one in which two axes are equal and one spatial component of the axial vector torsion does not vanish, giving rise to a universe that is spatially shaped as an ellipsoid of rotation revolving about the only axis along which the spin density is not equal to zero. It is also worth noticing that for the specific form of the Dirac field we have chosen, although the Dirac field is massive, still we have the conservation of the axial current; on the other hand it is widely known that quantum corrections might produce additional terms for which the axial current is not conserved anymore, if the Dirac field is charged or if additional terms in the dynamics are introduced, and then it would be interesting to see what are the effects that constraints such as \eqref{3.11} have on these partially conserved axial currents. However because in the present paper we do not deal with quantum corrections, nor with charged fields or extensions in the electrodynamics, addressing this issue would bring us far from the main aim of this work, and we refer the reader to \cite{m/3} and references therein for a discussion on this subject.
\\
The field equations have been worked out in both the Jordan and the Einstein frames, showing that the procedure to get solutions proposed in \cite{Saha1,Saha2} applies also here, independently of the given function $f(R)$. In particular, in the Einstein frame the Einstein-like equations result in general to be quadratically integrable.
\\
As it may have been expected, these models are intrinsically quite difficult to study in general; we have focused on some specific examples for which the field equations are considerably simplified: we have studied the model $f(R)=R^4$, where exact analytic solutions are found, although this model does not reproduce the physical content we expect; we have further studied the model $f(R)=R+\alpha R^2$, for which exact analytic solutions are very difficult to obtain; therefore, in this case, we have first worked in the Jordan frame, finding solutions that are analytic but approximated to the first order of $\alpha$, and then we have moved in the Einstein frame, employing numerical methods. In the case $f(R)=R^4$, we have found that the universe does not display isotropization, as it should, and there is no correct Einstein limit; in the $f(R)=R+\alpha R^2$ case, we have found that the results obtained in the two frames coincide, and they yield Einsteinization and isotropization, that is as the cosmic time goes by both the non-linearities of the $f(R)$ function and the anisotropies of the background tend to vanish. In both models however, no initial singularity for the volume of the universe has been found at all.
\\
Finally, we have proved that the proposed method for solving the field equations is consistent with the assignment of initial data, clarifying the role of the previously stated general conservation laws \cite{FV1} in preserving the Hamiltonian constraint.
\\
To summarize, we have found here three types of results: first of all, from the point of view of the behaviour of the universe, our results may help to solve one of the major conceptual problems of $f(R)$ theories related to the arbitrariness of the $f(R)$ function; indeed, as we have shown, a given $f(R)$ may either give rise to physics with no good behaviour or yield physics with good behaviour; thus although there is no theoretical reason to prefer a particular form for the $f(R)$ function instead of another, nevertheless there are phenomenological set-ups in which selection rules discriminating physical $f(R)$ function from infeasible ones may be established. Secondly, universes that have finite initial volume are possible, and therefore the gravitational singularity problem receives new fuel from the present discussion. Finally, the initial values problem is shown to be well formulated.
\\
The general outlook emerging from our study encourages to pursue the study of torsional $f(R)$ models, since their field equations and conservation laws are well defined and consistent with the assignment of initial data, while the presence of torsion can remove the singularity problem, and there are hopes that a possible specific $f(R)$ may be selected which will be able to describe what is missing in the evolution of the universe as we know it.
%%%%%%%%%%%%%%%%%%%%%%%%%%%%%%%%%%%%%%%%%%%%%%%%%%%%%%%%%%%%%%%%%%%%%%%%%%%%%%%%%%%%%%%%%%%%%%%%%%%

%%%%%%%%%%%%%%%%%%%%%%%%%%%%%%%%%%%%%%%%%%%%%%%%%%%%%%%%%%%%%%%%%%%%%%%%%%%%%%%%%%%%%%%%%%%%%%%%%%%
%%%%%%%%%%%%%%%%%%%%%%%%%%%%%%%%%%%%%%%%%%%%%%%%%%%%%%%%%%%%%%%%%%%%%%%%%%%%%%%%%%%%%%%%%%%%%%%%%%%
\end{document}